\documentclass[preprint,review,12pt, authoryear]{elsarticle}

\usepackage{amssymb}
\usepackage{amsthm}
\usepackage{amsmath}
\usepackage{xcolor}
\usepackage{algorithm}
\usepackage{algorithmic}

\usepackage{caption}

\usepackage{listings}

\journal{Computer methods and programs in Biomedicine}

\begin{document}

\begin{frontmatter}



\title{Bayesian inference for case-control point pattern data in spatial epidemiology with the \texttt{inlabru} R package}


\author[label1]{Francisco Palm\'i-Perales \corref{cor1}}
\author[label2]{Finn Lindgren}
\author[label3]{Virgilio G\'omez-Rubio}

\address[label1]{Departament of Statistics and Operational Research, Universidad de València. \\C/ Dr. Moliner, 50, 46100, Burjassot (Valencia, Spain)}
\address[label2]{School of Mathematics and Maxwell Institute for Mathematical Sciences, \\ University of Edinburgh, Edinburgh EH9 3FD, United Kingdom}
\address[label3]{Departamento de Matem\'aticas, Universidad de Castilla-La Mancha. \\ Av, Espa\~na s/n, 02071, Albacete (Spain)}

\cortext[cor1]{Francisco Palmí Perales (Corresponding author) \\ 
			   Department of Statistics and Operative Research, Universitat de València, \\
			   C/ Dr. Moliner, 50, 46100, Burjassot (Valencia, Spain)\\
			   Francisco.Palmi@uv.es \\
			   (+34) 963 54 43 08 }

\begin{abstract}
\textbf{Background and Objective:} 
The analysis of case-control point pattern data is an important problem
in spatial epidemiology. The spatial variation of cases if often compared
to that of a set of controls to assess spatial risk variation as well as the
detection of risk factors and exposure to putative pollution sources using spatial regression models. \\
\textbf{Methods:} The intensities of the point patterns of cases and controls are estimated using log-Gaussian Cox models, so that fixed and spatial random effects can be included. Bayesian inference is conducted via the integrated Nested Laplace approximation (INLA) method using the \texttt{inlabru} R package. In this way, potential risk factors can be assessed by including them as fixed effects while residual spatial variation is considered as a Gaussian process with Matérn covariance. In addition, exposure to pollution sources is modeled using different smooth terms.\\
\textbf{Results:} The proposed methods have been applied to the Chorley-Ribble dataset, that records the locations of lung and larynx cancer cases as well as the location of an disused old incinerator in the area of Lancashire (England, United Kingdom). Taking the locations of lung cancer as controls, the spatial variation of both types of cases has been estimated and the increase of larynx cases in the vicinity of the incinerator has been assessed. The results are similar to those found in the literature. \\
\textbf{Conclusions:} A framework for Bayesian analysis of multivariate case-control point patterns within an epidemiological framework has been presented. Models to assess spatial variation and the effect of risk factors and pollution sources can be fit with ease with the \texttt{inlabru} R package.\\
\end{abstract}




\end{frontmatter}
\pagebreak



\section{Introduction}
\label{intro}

\medskip
Spatial statistics has been an essential tool for epidemiologists and public health scientists in recent decades. Incorporating the geographic locations of cases enables researchers to discern the disease pattern across the study area. Moreover, taking into account additional information helps when assessing the influence of risk factors on the disease pattern. Integrating all information available from different sources will allow researchers to build more appropriate models to study the spatial variation of diseases and associated risk factors. A recent review of statistical methods for spatial epidemiology is available in \citet{HandSpEpi}.

\medskip
It is often the case that the actual locations of the different events, usually cases and controls, are known and the aim of the epidemiological study is mainly to estimate the spatial patterns over the study region. In this case, a point pattern analysis shall be considered \citep{illian2008ppp, diggle2013ppp, baddeley2015ppp}.

\medskip
From a statistical point of view, when the events (points) can be classified into different groups, the resulting structure is a multivariate (or marked) point pattern. The most common example in spatial epidemiology is the analysis of the cases of different diseases which are compared to a set of controls or different types of a specific disease over a common study area. For example, \citet{Diggle2005bovine} study the spatial variation of cases of different strains of bovine tuberculosis in Cornwall (United Kingdom). 

\medskip
Point processes (i.e., the underlying mechanism that generates a point pattern) have often been developed by the motivation of analysing epidemiological data \citep{DiggleVirgilio2007} and several authors have shown that Cox processes \citep[specifically, log-Gaussian Cox processes,][]{LogGaussianCox} provide an appropriate framework for modelling point processes that are environmentally driven
\citep{liang2008analysis, diggleMoraga2013, Waagepetersen2016}.

\medskip
The analysis of multivariate point patterns in case-control studies using log-Gaussian Cox process have been explored in different ways. In this paper, we will focus on those based on Bayesian inference. \citet{diggleMoraga2013} illustrate the wide use of the log-Gaussian Cox processes with four different applications in a Bayesian framework using Markov chain Monte Carlo (MCMC) methods \citep{MCMC}. Alternatively, \citet{Illian2012} describe a toolbox for analysing point patterns using the integrated nested Laplace approximation \citep[INLA,][]{INLA}. Furthermore, \citet{Simpson2016} propose a new and computationally convenient method to perform inference on log-Gaussian Cox processes by approximating the likelihood using a continuously specified Gaussian random field. Recently, \citet{krainski2018advanced} illustrate some advanced features of the analysis of multivariate point patterns using INLA.

\medskip
Regarding other approaches to the analysis of multivariate point patterns, \citet{Virgilio2015Tornados} propose a novel framework in which a spatial component is shared among all patterns and a specific spatial component is considered for each point process. \citet{PalmiPerales2019Biometrical} and \citet{RJournal2023} study the spatial variation of a set of controls and the cases of three different types of cancer so that the spatial variation of the controls acts as a baseline of the spatial variation of the cancer cases to assess spatial risk variation as well as exposure to different putative pollution sources.

\medskip 
The main aim on this work is to present a detailed procedure and software of how to analyse point patterns in an epidemiological context using spatial models and Bayesian inference. For this, the \texttt{inlabru} R package \citep{inlabru} will be used because of two main reasons. First of all, it offers a simple and friendly coding environment. Secondly, it relies on the INLA method and the \texttt{R-INLA} R package to estimate log-Gaussian Cox models efficiently by using recent methodological developments \citep[SPDE,][]{SPDE, Simpson2016}. Thus, researchers which are not experts in programming will be able to analyse the data coming from their studies benefiting from recent statistical developments using a friendly software environment.

\medskip
The structure of this paper is as follows. Section \ref{sec:mpp} introduces the main methodology to fit spatial regression models to multivariate point patterns within an epidemiological framework. Next, Section \ref{sec:mfitting} describes model fitting. In Section \ref{sec:example}, the analysis of the Chorley-Ribble dataset \citep{Diggle1990splancs} is conducted using the proposed methodology and software. Lastly, a discussion and some concluding remarks are provided in Section \ref{sec:disc}.




\section{Multivariate point pattern analysis}
\label{sec:mpp}

\medskip
Given a bounded study region $D \subseteq \mathbb{R}^2$, a point pattern is a set of points in $D$ (i.e., $\{ x_i \}_{i=1}^n$). These points are the realization of a point process \citep[see, for example, chaper 4 of][]{diggle2013ppp}, which can be regarded as a stochastic process which determines the location of the different events. For instance, in epidemiology, an typical example of a point pattern is the location of where people that suffer from a specific disease live. 

\medskip
Multivariate point pattern datasets contain the locations of events of different classes or groups. Apart from the coordinates of the locations of the different events and the marks that determine to which point pattern each event belongs, extra information could be found in these datasets, such as environmental or socio-economic factors. The availability of these information influences the definition of the model. 

\medskip
The homogeneous Poisson point process is the simplest point process model \citep{illian2008ppp}. This point process is characterised by two fundamental properties: a) the number of points in a bounded region follows a Poisson distribution with a mean equal to the product between a constant $\lambda$ (called the intensity) and the area of the study region, and, b) the independence of points which belong to different disjoint regions. Parameter $\lambda$ is denoted the intensity of the point process and it represents the average number of point per unit area. In case $\lambda$ is not assumed to be constant, the resultant point process is called inhomogeneous Poisson process. In this case, the intensity depends on the location $x$ and it is denoted by $\lambda(x)$. Note that a convenient
way of modeling this non-constant intensity is a log-Gaussian Cox process \citep{LogGaussianCox}, as described below.

\medskip
Regarding modelling point patterns, let us consider a set of $P+1$ patterns,
so that there is a set of controls and $P$ patterns of cases of different
diseases. The way in which the intensity of the $i$-th point pattern, $\lambda_i(x)$, is estimated using a log-Gaussian Cox process is to define the logarithm of the intensity as follows:

$$
\log{(\lambda_i(x))} = \alpha+ \ldots + S(x) ;\,\,\, \forall i=0,\ldots, P \,\,\, x\in D .
$$

\noindent
Here, $\alpha$ is an intercept, '$\ldots$' represents a list of additive terms (e.g., fixed and random effects) and $S(x)$ is a spatial random effect that captures any residual spatial variation. Note also that intensity $\lambda_0(x)$ represents that of the controls.

\subsection{Modeling spatial variation}

Several scenarios with different structures can be proposed to estimate the intensity of the different point patterns. No covariates or other terms (with the exception of a spatial term) will be included now for simplicity. For instance, a baseline spatial model can be established considering the intensity of each point pattern separately:

$$
\log(\lambda_i(x))= \alpha_i + S_i(x); \, \, \, \forall i=1, \ldots, P
$$
\noindent
where $\alpha_i$ is the intercept of the $i$-th point pattern and $S_i(x)$ is the spatial continuous effect. Albeit this approach could simply estimate the spatial variation of each point pattern, this model assumes that the different point patterns are independent from each other (this is equivalent to analyse each point pattern separately). Thus, this model does not consider any extra information which could affect the point pattern. Consequently, unless a more complex model is considered, only simple comparisons (which do not consider the interactions between each point pattern) can be made between the different point patterns. Hence, more complex models scenarios should be developed to tackle the analysis of a case-control scenario.

All spatial random effects in this paper will be defined using a Gaussian
process with zero mean and covariance using a Matérn covariance function. This
function defines the covariance between the spatial effects at any two points $x_k$ and $x_l$ whose distance is $d_{k,l}$ as

$$
Cov(S(x_k), S(x_l))=\sigma^2\frac{2^{1-\nu}}{\Gamma(\nu)}\left(\kappa d_{k,l}\right)^{\nu}K_{\nu}\left(\kappa d_{k,l}\right).
$$
\noindent
Here, $\sigma^2$ is a variance parameter, $\nu$ is a positive smoothness parameter and
$\kappa$ is a positive scale parameter that controls how the covariance decays with the distance. $\Gamma(\cdot)$ is the gamma function and $K_{\nu}(\cdot)$ is the is the modified Bessel function of the second kind. 

\medskip
Note that this type of spatial random effects can be effectively fitted with INLA as described in Section
\ref{subsec:SPDE}. Also, note that the value of $\nu$ will be fixed \citep[see, for example,
Section 2.1.3 in][for details]{krainski2018advanced} and that instead of
$\kappa$ the nominal range $r$ will be used. This is defined as
$\sqrt{8\nu}/\kappa$ and it is the distance for which the correlation between
any two points is of about 0.14, i.e., if two observations have a distance larger
than the nominal range the spatial correlation will be small or
negligible.

\subsection{Assessing spatial variation}
\label{sec2:sub2}

\medskip
Regarding a multivariate point pattern analysis in a case-control study, it is always advised to model cases and controls jointly \citep{DiggleVirgilio2007}. The intensity of the cases reflects both variation in disease risk and variation in the intensity of the population at risk. The inclusion of the set of controls allows us to adjust for the spatial distribution of the population at risk and to identify risk factors associated with the disease. Therefore, in a case-control study it is appropriate to consider the estimation of both intensities, cases and controls, and compare them. 

\medskip
Controls will provide a baseline of the spatial distribution of the population at risk and they are assumed to be a realization of an inhomogeneous Poisson process with intensity $\lambda_0(x)$ \citep{diggle2013ppp}. Cases will be considered as realizations of different inhomogeneous point processes with their corresponding intensities, which may differ from that of the controls. When risk factors do not influence the occurrence of cases, their spatial pattern will closely resemble that of the controls. Conversely, if the spatial pattern of cases deviates from that of the controls, it may be an indication of discrepancies cause by relevant risk factors.

\medskip
Following the notation in \citet{PalmiPerales2019Biometrical}, a convenient way of modeling the intensities of the controls and $P$ groups of cases can be:

\begin{align*}
\log(\lambda_0(x)) &= \alpha_0 + S_0(x),\\
\log(\lambda_i(x)) &= \alpha_i + S_0(x)+ S_i(x),\, \, \, \forall i = 1, \ldots, P, 
\end{align*}

\noindent
where $\alpha_0$ and $\alpha_i$ are the intercept of the controls and the $i$-the group of cases, respectively. $S_0(x)$ is a spatial effect shared among all point processes, while $S_i(x)$ is another spatial effect specific of the $i$-th point process. 

\medskip
In order to compare the spatial patterns of the different spatial patterns, \citet{Diggle2005bovine} proposed to compute the ratio of the estimated intensities of cases and controls. In our case, if the estimation is done in the log-scale, it becomes:

\begin{equation}
\log(\frac{\lambda_i(x)}{\lambda_0(x)})= \log(\lambda_i(x)) - \log(\lambda_0(x))=$$
$$ (\alpha_i + S_0(x)+ S_i(x)) - (\alpha_0 + S_0(x)) = (\alpha_i - \alpha_0) + S_i(x)
\label{eq:logrisk}
\end{equation}
\noindent
Hence, the comparison of each group of cases and controls relies on the specific spatial term $S_i(x)$. Thus, this structure allows us to investigate whether there is any difference between each group of cases and the controls by studying the estimate of the case-specific spatial effect.

\subsection{Detection of high risk areas}
\label{sec2:sub3}

\medskip
High risk areas are parts of the study region where the number of cases is higher than expected according to the distribution of the controls.  Detecting this areas is essential in case-control studies. Following the previous approach, the shared spatial effect, $S_0(x)$, accounts for the spatial distribution of the population at risk distribution whereas the specific effect, $S_i(x)$, captures departures from the general spatial trend for each group of cases.

\medskip
As it can be seen in Equation~(\ref{eq:logrisk}), $\alpha_i - \alpha_0$ is a baseline of no risk. Departures
from this baseline in the spatial distribution of the cases will be picked up by  the specific spatial effect $S_i(x)$ (and, possibly, other terms included in the linear predictor when modeling the log-intensity of the cases).

\medskip
Specifically, if there is any factor which is affecting the spatial patterns of the cases, the specific effect should pick it up. Therefore, plotting the posterior mean of the spatial specific effect and analysing  95 \% credible intervals of its posterior distribution, it will be possible to detect any region with a increased risk in the study region. 

\subsection{Analysis of the effect of risk factors}
\label{sec2:sub4}

\medskip
As stated above, multivariate point pattern datasets may contain additional information about variables which could be convenient to account for. Previous models can be modified in order to consider additional factors included as covariates: 

\begin{align*}
\log(\lambda_0(x)) &= \alpha_0 + S_0(x)\\
\log(\lambda_i(x)) &= \alpha_i + S_0(x)+ F_i(x),\,\,\, \forall i = 1, \ldots, P; 
\end{align*}
\noindent
where, as before, $\lambda_0(x)$ and $\lambda_i(x)$ are the intensities of the controls and the $i$-th group of cases (respectively), $\alpha_0$ and $\alpha_i$ are the intercepts and $S_0(x)$ is the spatial shared effect. Now, $F_i(x)$ corresponds to the covariate effect on disease $i$ which can be specified in different ways. For example, $F_i(x)$ could be included as a simple linear effect on a covariate $z_x$ (i.e. $F_i(x) = \beta z_x$), as a linear term depending on a spatially varying covariate $z(x)$ (i.e. $F_i(x) = \beta z(x)$) or as a general smooth term (for example, distance to a pollution source).  When several covariates are available, such as a socio-economic or environmental information, they can be included  using $p$ different terms, e.g., $F_i^{(1)}(x) + \ldots + F^{(p)}_i(x)$.

\medskip
A particular case of exposure to risk factors appears when dealing with exposure to pollution sources such as industrial facilities, incinerators, etc. Each of these sources may increase the number of cases in its vicinity. In other words, the spatial distribution of the cases may be influenced by the distance to this source. This exposure effect can be modelled following different approaches.

\medskip
\citet{PalmiPerales2019Biometrical} consider the models proposed above and propose three different ways of including exposure effects using the distance to the pollution source. The first option is to consider a linear effect on the distance. Alternatively, smooth terms on the distance can be proposed. Alternatively, a random walk of order two on the distance  would induce a smooth estimate of the effect. Finally, they also consider this effect as a Gaussian process in one dimension so that the effect can take a smooth form. This Gaussian process will be modeled to take zero mean and
variance defied by a Matérn covariance function in one dimension (i.e. using the
distance from a given point to the pollution source).

\section{Bayesian inference for model fitting}
\label{sec:mfitting}

\medskip
As stated, Bayesian inference will be used to estimate the model parameters and
other quantities of interest.  In Bayesian inference, the final objective is to
obtain the joint posterior distribution of the parameters of the different
models.  In particular, inference will be conducted using the integrated nested
Laplace approximation method \citep[INLA,][]{INLA}. INLA is particularly well suited
to deal with models that  can be expressed as a latent Gaussian Markov random
field and can exploit the particularities of these models for computational efficiency. 

\medskip
In a nutshell, INLA aims at estimating the posterior marginal distribution
of the latent effects and hyperparameters by using numerical integration
and the Laplace approximation. Hence, INLA is not based on costly
sampling methods such as Markov chain Monte Carlo algorithms.
\citet{NewINLA2023} provide a brief overview of the methodology and its applications and give a detailed description of the INLA algorithm.

\medskip
Once the posterior marginal distributions have been obtained, INLA is able to
compute summary statistics such as the posterior mean, 95\% credible intervals
and other quantities of interest. In addition, INLA can compute several model
assessment and choice criteria that can be handy when comparing different models.

\subsection{Estimation of the intensity}
\label{subsec:SPDE}

\medskip
When it comes to estimating the intensity of a point pattern different methods
can be followed \citep[see, for example,][]{baddeley2015ppp, illian2008ppp}.
The models presented above are defined as log-Gaussian Cox processes that
can include some fixed effects, smooth terms on the covariates and one
or several smooth terms which will be modeled as Gaussian processes with zero
mean and covariance defined by a Matérn covariance function.

\medskip
Estimation of the different spatial effects is conducted by expressing
them as the solution to an stochastic partial differential equation (SPDE),
as described in \cite{SPDE}.
In particular, the SPDE approximation for estimating a Gaussian process with Matérn covariance is based on solving a stochastic partial differential equation \citep[SPDE, ][]{SPDE} in a weak form. This approach allows the estimated spatial effect \( S(x) \) to be expressed as:

\[
S(x) = \sum_{k=1}^m \psi_k(x) w_k
\]
\noindent
In this representation, \(\psi_k(x)\) are basis functions, and \(w_k\) are Gaussian weights. Index \(k\) refers to a particular vertex in a triangulation (or mesh) that covers the study region. Each basis function \(\psi_k(x)\) is piecewise linear within the triangles of the mesh, equal to 1 at vertex \(k\) and 0 at all other vertices. This sparse representation is computationally efficient and practical for large-scale spatial data analysis.

The SPDE approach is particularly advantageous as it transforms the problem of working with dense covariance matrices into a problem involving sparse precision matrices. This is achieved by representing the spatial field through a finite element method (FEM) using a mesh, making the computations more tractable. When reporting results from this method, the nominal variance and nominal range are used to summarize the estimates of the spatial effects \( S_i(x) \) for \(i = 0, \ldots, P\).

For more details, the reader is referred to \cite{krainski2018advanced}, \cite{blangiardo2015spatial} and \cite{gomez2020bayesian}. It is worth mentioning that we need to properly build the mesh used by the SPDE approximation because its choice could affect the numerical accuracy of the estimation. Some advice on this point can be found in Chapter 2 of \citet{krainski2018advanced}.

\subsection{Prior distribution choice}
\label{sec3:PCpriors}

\medskip
Prior distributions are an essential element of Bayesian analysis and have to be properly chosen. The implementation in \texttt{R-INLA} and \texttt{inlabru} usually relies on some default prior choices, therefore if a prior distribution is not specified, these functions will automatically apply the default options. However, these default priors may not be the appropriate for every context. 

\medskip
A wide range of prior distributions is also available in \texttt{R-INLA}. Furthermore, this package provides several ways to define a different prior distribution for the hyperparameters. The reader is referred to Chapter 5 of \citet{gomez2020bayesian} and to \citet{krainski2018advanced} where different ways of how to define a prior in INLA are detailed. 

\medskip
Regarding the priors chosen for the analysis, default priors will be used for
the intercept and coefficients of the fixed effects. For the intercept, an
improper Gaussian prior with zero mean and zero precision is used. For the
coefficients of the fixed effects, the default prior is a Gaussian distribution
with zero mean and precision of 1000.

\medskip
When analysing log-Gaussian Cox processes the prior choice of the spatial hyperparameters has also to be done carefully. 
For the parameters of the spatial effects, penalized complexity priors \citep[PC-priors, ][]{PCpriors} will be used. The building idea of this prior distributions follows the principle of parsimony by penalising the distance from a base model. Furthermore, they are set using intuitive probability statements.

\medskip
Although it is difficult to provide a general advice on the choice of priors
for the models discussed in this paper, we would like to shed a bit of light on the choice of the parameters of the prior distributions for the nominal range and the nominal variance of a spatial effect. 
\citet{Simpson2019Careful} describe the details of how to choose prior distributions for the hyperparamenters of a spatial effect in the analysis of log-Gaussian Cox processes in a regular grid.

\medskip
Informally, the range represents the distance limit beyond which the data are no longer correlated. Therefore, a reasonable approach to set a prior is to pick half the maximum distance (``diameter") inside of the study region ($max_d$), approximately, and specify a low probability that the range will be bigger than this distance, e.g.: 

$$ P(r < 0.5 \cdot max_d ) = 0.99 $$

\medskip
In the case of the standard deviation, our proposal follows the ideas in \citet{Simpson2019Careful}. Specifically, an upper limit for the variation of the intensity ($U_{\alpha}$) is considered: 

$$ P(\sigma > U_{\alpha}) = 0.01 $$

\medskip
The value of this upper limit will depend on each problem as it measures how the intensity will change across the study region. The value of $U_{\alpha}$ can
be set by thinking on how we expect the intensity to vary across the study region.

\medskip
Finally, it is also highly recommended to perform a sensitivity analysis in order to assess the effect of the prior choose. 

\subsection{Software}

\medskip
The INLA method is implemented in the \texttt{R-INLA} package while
\texttt{inlabru} is an R package designed to facilitate spatial and spatio-temporal modeling using INLA. It is particularly useful for Bayesian inference in highly parameterized models, offering a flexible framework for fitting models to data that might be spatially or temporally structured. It provides an user-friendly interface for defining complex models using familiar formula syntax, making it easier to fit models compared to using \texttt{R-INLA} directly. \texttt{inlabru} is particularly effective for point process models and integrates well with other R packages for spatial data handling and visualization.

\medskip
When developing the example, \texttt{sf} \citep{sf} and \texttt{terra} \citep{terra} R packages have been used for data management and visualization.  Furthermore, the maps included in this paper have been created with R packages \texttt{ggplot2} \citep{ggplot2} and \texttt{tmap} \citep{tmap}.

\section{Example}
\label{sec:example}

\medskip
In this section, a multivariate point pattern dataset is going to be analysed in order to exemplify the methodology presented in this paper using the \texttt{inlabru} (version 2.12.0) and \texttt{INLA} (version 24.06.27) R packages.
The chosen dataset is the \texttt{Southlancs} dataset, which is available in the \texttt{splancs} R package \citep{splancs}. This dataset contains the locations of the cases of larynx and lung cancer and the location of an disused incinerator which was closed in 1984. All the cases are located in the Chorley-Ribble area of Lancashire, England. Coordinates are in kilometres, and the resolution is 100 metres (0.1 km). The choice of this dataset has been motivated by the fact that it has been widely analysed \citep{diggle1990southlancs, Diggle1990splancs, DiggleRowlingson1994splancs} and it is accessible for anyone who wants to strictly reproduce this example. 

\medskip
Next, a multivariate analysis of this dataset will be performed step by step considering the lung cancer cases as the baseline point pattern which is usually assumed to be the controls \citep{Diggle1990splancs, Diggle1994}. 
Algorithm \ref{alg:inlabru} illustrates the different steps to conduct a
multivariate analysis of point patterns with \texttt{inlabru}.

\begin{algorithm}
\caption{Analysing point patterns data with \texttt{inlabru}}\label{alg:inlabru}
\begin{algorithmic}[1]
\STATE Load the required data.
\STATE Build the mesh.
\STATE Build the SPDE basis function.
\STATE Define the model likelihoods and components.
\STATE Fit the model.
\STATE Extract the results.
\STATE Display and analyse the results.
\end{algorithmic}
\end{algorithm}

\begin{figure}
\centering
\includegraphics[width = 6.5 cm]{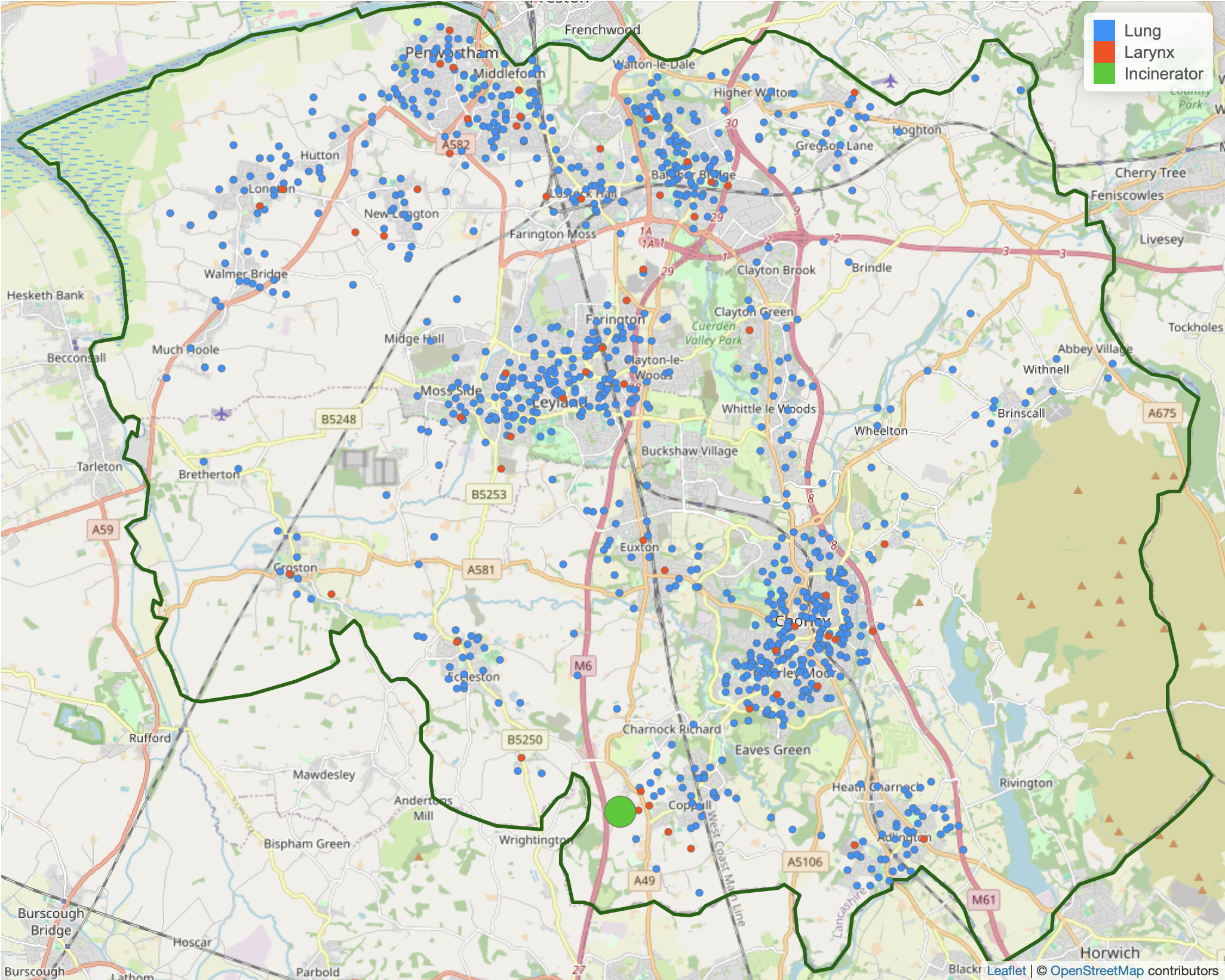}
\includegraphics[width = 5.5 cm]{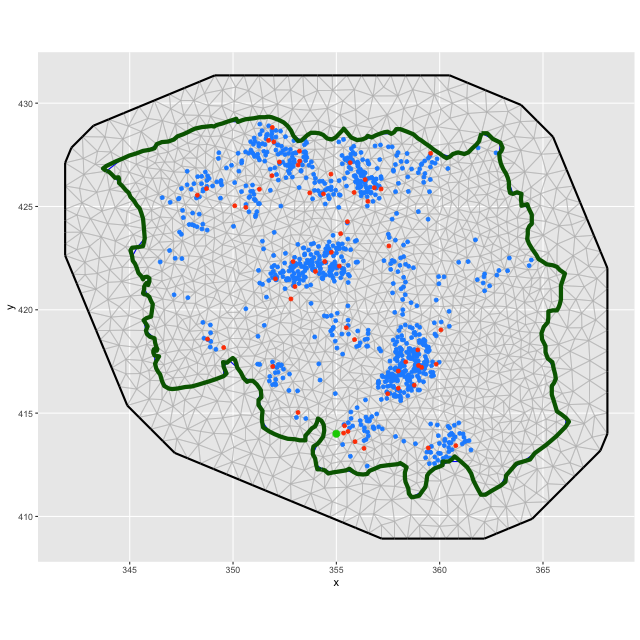}
\caption{Location of the lung and larynx cancer cases among the Chorley and South Ribble Health Authority of Lancashire (left) and mesh used by the SPDE approach (right).}
\label{fig1}
\end{figure}

\medskip
The left plot in 
Figure \ref{fig1} shows the location of lung cancer cases (blue), the location of larynx cancer cases (purple), the location of the disused incinerator (red) and the boundary of the study region in the Chorley and South Ribble Health Authority of Lancashire (England, U.K.).

\subsection{Estimation of the spatial variation}
\label{sec3:sub1}

\medskip
The example starts with a univariate analysis of each point pattern separately. For this reason, the results from the univariate analyses can be seen as baseline results in order to compare them with the results from more complex models. 

\medskip
The right image in Figure \ref{fig1} shows the mesh which has been used for analysing \texttt{southlancs} data is displayed. Building the mesh will be the first step to fit a SPDE model. Mesh design can influence the computation of the results, therefore, it has to be done following some basic concepts. First, the triangles should be as regular as possible in size and shape in order to maintain a constant estimation over the whole region. It is also essential to avoid the boundary effect by setting an outer part of the mesh with a significant distance from the boundary of the study region \citep{lindgren2012continuous}. \cite{krainski2018advanced} provide some relevant details to take into account when dealing with the mesh definition. Furthermore, the mesh should be small enough as to pick short distance effects as well as the spatial variation of the effect of the covariates.

\medskip
The recommended function to create the mesh is \texttt{fm\_mesh\_2d\_inla()}.
As an example, here we show the code used to build the mesh we have used in this study: 

\begin{verbatim}
mesh <- fm_mesh_2d_inla(
  boundary = list(bdy_sf, NULL),
  cutoff = 0.4,
  max.edge = c(0.6, 1.2),
  min.angle = 27,
  offset = c(0.5, 2),
  n = c(16, 16)
)
\end{verbatim}

\medskip
The \texttt{boundary} argument is used to specify the inner and/or the outer
boundaries; here, \verb+bdy_sf+ is the object that stores the boundary of the study region. The \texttt{cutoff} argument allows you to state a minimum distance
between two nodes (triangle vertices). The \texttt{max.edge} and the
\texttt{min.angle} arguments establish the maximum length of a triangle edge
and the minimum inner angle. The \texttt{n} argument specifies the number of
initial nodes, while the \texttt{offset} argument determine the distance of the
outer boundary. 

\medskip Furthermore, a tool for building meshes is available by calling 
function \texttt{meshbuilder()},
which opens a Shiny interface \citep{shiny} that allows  to choose convenient
values for the arguments of the \texttt{fm\_mesh\_2d\_inla()} function in an
interactive environment. This can be very useful to find the optimal
values of the arguments easily.

\medskip
Once the mesh is defined, we should build the basis functions to estimate the spatial effect using the SPDE approach. At the same time, the prior distributions of the spatial hyperparameters (the range and the nominal standard deviation) are specified. When it comes to specifying the prior distributions, several options can be considered. In this analysis, PC-priors \citep{PCpriors} will be set using the \texttt{inla.spde2.pcmatern()} function. Specifically, the prior distributions set on the nominal range ($r$) and the nominal standard deviation ($\sigma$) are: 

$$ P(r < 10) = 0.99$$
$$ P(\sigma > 1) = 0.01$$
\noindent
Note that $\sigma$ is the standard deviation of a Gaussian random effect in the log-scale. Hence, values larger than one will imply very large variation of the intensity, which is not reasonable for the current analysis.

These prior distributions are set specifically for this analysis. A sensitivity analysis (available as a supplementary material) has also been performed in order to assess the effect of the prior choice for the nominal standard deviation. Some comments about the choice of the parameters of the PC-priors have been discussed in Section \ref{sec3:PCpriors}. Values have been chosen applying this criterion to this dataset. Here we can see the related lines on the code: 

\begin{verbatim}
pcmatern <- inla.spde2.pcmatern(
  mesh = mesh, alpha = 2,
  prior.range = c(10, 0.99),
  prior.sigma = c(1, 0.01)
)
\end{verbatim}
\noindent
Note that argument \verb+alpha = 2+ is used to set parameter $\nu$ equal to
1 in the definition of the Matérn covariance (because $\nu$ is equal to 
\verb+alpha+ minus the dimension divided by 2 \cite{SPDE}).

\medskip
Once the setting to estimate the spatial effects has been defined, 
the model components will be defined to estimate the different intensities.
First of all, univariate models are defined as:

$$
\log(\lambda_i(x))= \alpha_i + S_i(x); \, \, \, \forall i=0,1.
$$

This model is defined using an R formula as follows:

\begin{verbatim}
cmp <- geometry ~
   Intercept(1) + spatialspde(geometry, model = pcmatern)
\end{verbatim}

\medskip
The \texttt{geometry} term on the left hand side represents the spatial coordinates of the points. The intercept of the model ($\alpha_i$) is represented by \texttt{Intercept(1)}, then the spatial effect ($S_i(x)$) is denoted by the last 
component \texttt{spatialspde(geometry, model = pcmatern)}, which takes two arguments:
the coordinates of the points (\texttt{geometry}) and a model specification for the SPDE (variable \verb+pcmatern+ defined above).

\medskip
Although the spatial term is referred to as \texttt{spatialspde} in the previous code lines, this specific name is not mandatory. Any name may be used, provided it is consistently applied throughout the code. However, the structure within the parentheses must remain consistent. The first argument specifies the varying spatial field (in our case, over the coordinates of the point pattern), and the second argument, \texttt{model}, defines the correlation function (in our case, the SPDE model defined previously).

\medskip
At that point, we can fit the model using the \texttt{lgcp()} function of the \texttt{inlabru} package as follows: 

\begin{verbatim}
fit.lun <- lgcp(
  components = cmp,
  data = sl.lun,
  domain = list(geometry = mesh),
  samplers = bdy_sf
)
\end{verbatim}
\noindent
Note that this model has used dataset \verb+sl.lun+ that includes the locations of the cases of lung cancer. A similar model could be fit to the larynx cancer cases by replacing the value of argument \verb+data+ accordingly.

\begin{figure}
\includegraphics[width = 7.05 cm]{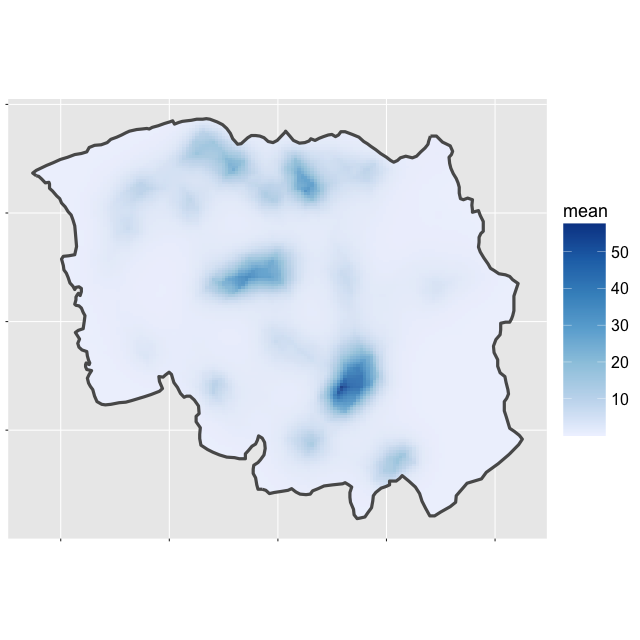}
\includegraphics[width = 7.05 cm]{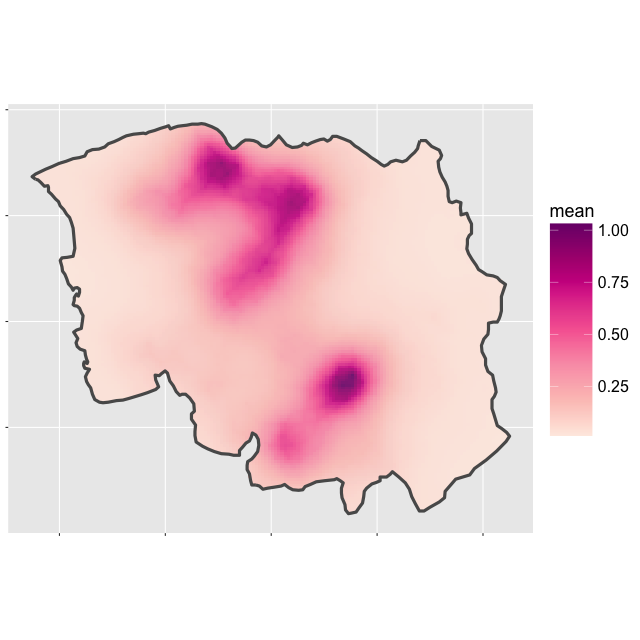}
\caption{Estimated intensity of the lung cancer cases (left) and the larynx cancer cases (right) been estimated separately. Note that the scales strongly differ.}
\label{fig2}
\end{figure}

\medskip
Once the model is fit, the estimation of the intensities can be displayed. Figure \ref{fig2} shows the estimates (posterior means) of the intensity of both lung (left) and larynx (right) cancer cases. These results have been obtained performing two univariate analysis, one for each group of cancer cases. In addition, the differences of the scales have to be taken into account when comparing the estimates as there are 58 larynx cancer cases and 978 lung cancer cases. 

\subsection{Comparing cases and controls}
\label{sec3:sub2}

\medskip
Although a univariate analysis is a reasonable starting point, a case-control study needs a deeper analysis through a more sophisticated approach. A logical next step is to model both lung and larynx cancer cases together. As it has been done in previous works \citep{Diggle1990splancs, Diggle1994}, lung cancer cases will be considered  as the controls in this analysis. This will allow us to compare the spatial distribution of the cases and the controls as described in Section \ref{sec2:sub2} and to detect areas of high risk in the study region as in Section \ref{sec2:sub3}. From now on, controls will refer to the cases of
lung cancer while cases will refer to the cases of larynx cancer.

\medskip
Firstly, we will consider a baseline model with only a spatial term $S_0(x)$ shared between both groups which will capture the spatial variation of the intensity of both point patterns as follows: 

\begin{align*}
\log(\lambda_{0}(x)) &= \alpha_{0} + S_0(x),\\
\log(\lambda_{1}(x)) &= \alpha_{1} + S_0(x) \, \, \, \forall x \in D; 
\end{align*}

\noindent
where $\lambda_{0}(x)$ and $\lambda_{1}(x)$ are the intensities of the controls and the cases, respectively. Furthermore, $\alpha_{0}$ and $\alpha_{1}$ represent the intercept of the controls and cases, respectively. 

\medskip
Next, a spatial cases-specific effect $S_{1}(x)$ will be added as discussed in Section \ref{sec2:sub2}. This will lead to a model with two likelihoods, one for each point pattern. Note that the $S_0(x)$ term will be shared between both likelihoods.

\begin{align*}
\log(\lambda_{0}(x)) &= \alpha_{0} + S_0(x)\\
\log(\lambda_{1}(x)) &= \alpha_{1} + S_0(x)+ S_{1}(x)\, \, \, \forall x \in D. 
\end{align*}
\noindent
Both spatial effects will be fit using the same mesh and priors as in the univariate models but different sets of parameters will be estimated now for each spatial term. 

Following this, the model components need to be specified. In this case, an intercept for each point pattern is set (named \texttt{Inter.con(1)} and \texttt{Inter.lar(1)}, respectively) and the spatial shared effect $S_0(x)$ (as component \texttt{sharedspde}). The \verb+-1+ term is added in order to remove the default intercept in the model. 
The code to define this is:

\begin{verbatim}
cmp <- coordinates ~ -1 + Inter.con(1) + Inter.lar(1) +
  sharedspde(coordinates, model = pcmatern) 
\end{verbatim}

As we are considering two likelihoods, they are specified using the components defined above and the \texttt{like()} function as follows: 

\begin{verbatim}
# Likelihood for the controls (lung cancer)
con.lik <- like(
  family = "cp",
  formula = geometry ~ Inter.con + sharedspde,
  samplers = bdy_sf,
  data = sl.lun,
  domain = list(geometry = mesh)
)

# Likelihood for the cases (larynx cancer)
lar.lik <- like(
  family = "cp",
  formula = geometry ~ Inter.lar + sharedspde,
  samplers = bdy_sf,
  data = sl.lar,
  domain = list(geometry = mesh)
)
\end{verbatim}
\noindent
\medskip
Here, argument (\texttt{family = "cp"}) indicates that we are adjusting a Cox process, then we specify the structure of each likelihood. Once the likelihoods are defined, the model can be fit using the \texttt{bru()} function: 

\begin{verbatim}
fit <- bru(cmp, con.lik, lar.lik)
\end{verbatim}

\medskip
As before, once the model is fit, we are able to display and summarize the intensity or any component in the model such as the spatial effect. 

\medskip
The model to be fit will include the spatial specific effect for the cancer, $S_1(x)$. This will allow us to assess differences between the intensities of cases and controls. Furthermore, spatial effect $S_1(x)$ will potentially show the areas with high risk.

\medskip
The components and the likelihood definitions will be the only lines modified by including this specific spatial term which has been called \texttt{larspde} in the code: 

\begin{verbatim}
cmp <- geometry ~ -1 + Inter.con(1) + Inter.lar(1) +
  sharedspde(geometry, model = pcmatern) +
  larspde(geometry, model = pcmatern)
\end{verbatim}
\noindent
Note that component \verb+larspde+ is a spatial random effect defined similarly as the other spatial effect but it will only be included when defining the 
components of the likelihood of the cancer cases.

\begin{verbatim}
# Likelihood of the controls (lung cancer)
con.lik <- like(
  family = "cp",
  formula = geometry ~ Inter.con + sharedspde,
  samplers = bdy_sf,
  data = sl.lun,
  domain = list(geometry = mesh)
)

# Likelihood of the cases (larynx cancer)
lar.lik <- like(
  family = "cp",
  formula = geometry ~ Inter.lar + sharedspde + larspde,
  samplers = bdy_sf,
  data = sl.lar,
  domain = list(geometry = mesh)
)
\end{verbatim}

\begin{figure}[h!]
\includegraphics[width = 7.50 cm]{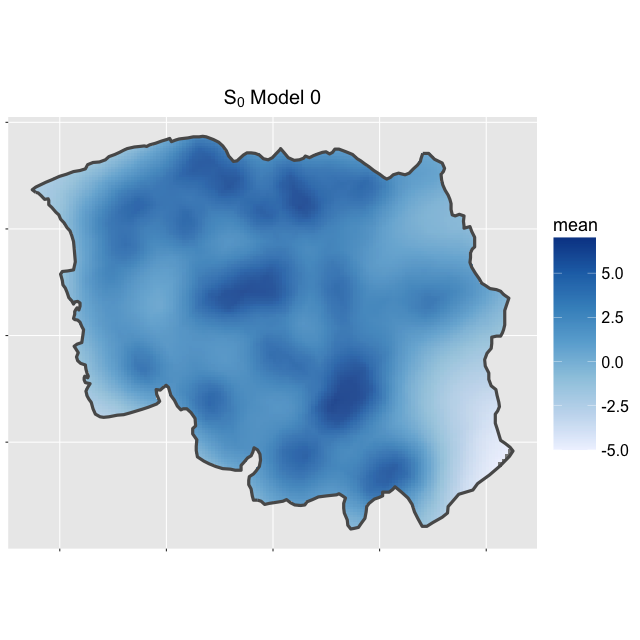}
\hspace{7.5 cm }\\
\includegraphics[width = 7.50 cm]{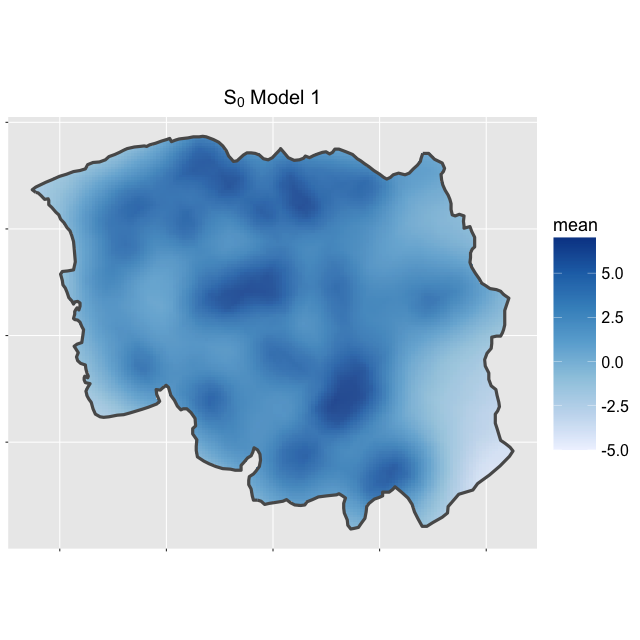}
\includegraphics[width = 7.50 cm]{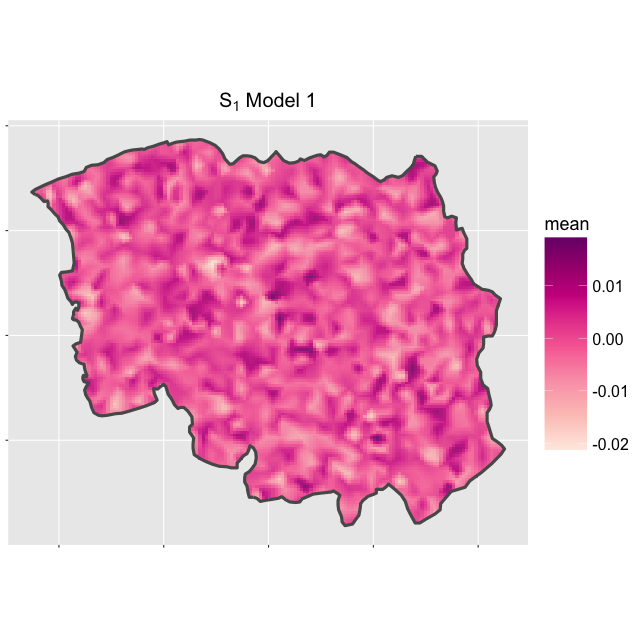}
\caption{Estimation of the posterior mean of the spatial common effect for model 0 (only shared spatial effect, top left), for model 1 (shared and specific spatial effects, bottom left) and the estimation of the posterior mean of the spatial specific effect of the larynx cancer cases (bottom right).   }
\label{fig3}
\end{figure}

\medskip
The estimates of the posterior mean of the common and specific spatial effects are shown in Figure \ref{fig3}. As it can be seen, there is a huge heterogeneity in the shared spatial effect which is most probably due to the urban concentration of the population \citep{Diggle1990splancs}. Additionally, the scale of the cases-specific spatial random effect is very small compared to the scale of the shared spatial random effect. This suggests a similar spatial pattern between cases and controls, which has been mainly captured by the spatial shared effect. Furthermore, the estimates of the shared spatial random effect in both models are very similar. This also suggests a very similar spatial pattern for both cases and controls because including the cases-specific spatial random effects does not modify the shared one in the second model.

\subsection{Exposure to pollution sources}

\medskip
The Chorley-Ribble dataset also contains the location of an disused incinerator which could be influencing the appearance of cases. Therefore, it is convenient to, at least, consider the effect of the incinerator on the intensity of the cases in the model structure. Therefore, three different alternatives of modelling the effect of the incinerator on the cases distribution are considered following the approaches discussed in Section \ref{sec2:sub4}.

\medskip
A simple option is to compute the distance between the incinerator location and that of every case and include the effect as a typical linear term. This will be specified in the components using \texttt{model = "linear"}: 

\begin{verbatim}
cmp <- geometry ~ -1 + Inter.con(1) + Inter.lar(1) +
  sharedspde(geometry, model = pcmatern) +
  larspde(geometry, model = pcmatern) +
  dist(dist_terra, model = "linear")
\end{verbatim}

Note that the first argument in component \texttt{dist} is defined as
\texttt{dist\_terra}. This is a shortcut to pass a spatial covariate
in a raster format so that \texttt{inlabru} automatically obtains the right value of the covariate according to the coordinates of any given point. This simplifies the code and avoids users from computing these values themselves.

Next, the two likelihoods are defined as:

\begin{verbatim}
# Likelihood for the controls (lung cancer)
con.lik <- like(
  family = "cp",
  formula = geometry ~ Inter.con + sharedspde,
  samplers = bdy_sf,
  data = sl.lun,
  domain = list(geometry = mesh)
)

# Likelihood for the cases (larynx cancer)
lar.lik <- like(
  family = "cp",
  formula = geometry ~ Inter.lar + sharedspde + larspde + dist,
  samplers = bdy_sf,
  data = sl.lar,
  domain = list(geometry = mesh)
)
\end{verbatim}

\medskip
In addition to considering a linear term on the distance to the incinerator,
other non-linear smooth terms can be included based on distance using a 
SPDE in one dimension or a 
random walk or a Gaussian process in one dimension (which can also be estimated using the SPDE approach). Next, a Gaussian process of order one and a random walk of order two will be considered to model the effect of the distance. 

\medskip
In both cases, the definition of the Gaussian process in one dimension requires a one dimensional mesh that can be created with the \verb+fm_mesh_1d()+ function as follows:

\begin{verbatim}
#Build a 1 dimension mesh for exposure effect 
mesh_cov <- fm_mesh_1d(
  seq(0, ceiling(minmax(dist_terra, TRUE)[2]), length = 20),
  degree = 2,
  boundary = c("free", "free"))
\end{verbatim}

\medskip
This mesh is in turn passed on to function \texttt{inla.spde2.matern()} to define the SPDE random effect to be used to define one of the components in the model:

\begin{verbatim}
#Create the basis of the covariate 1 dimension SPDE effect
spde_cov <- inla.spde2.pcmatern(
  mesh = mesh_cov,
  alpha = 2,
  constr = TRUE,
  prior.range = c(10 / 5, 0.99),
  prior.sigma = c(1, 0.01))
\end{verbatim}

\medskip
Note that here the prior is set so that the range is smaller than in the spatial SPDE random effects to assume that effects are localized and that the standard deviation is set to be a bit larger to allow for larger and local variation.

\medskip
Finally, the random walk of order two can be approximated using the SPDE approach by following the results in \citet{SPDE} by setting the range to a large fixed value. In the next code, the second argument of the prior on the range is set to \verb+NA+ so that it is fixed to a value equal to the first argument (10, in our case).

\begin{verbatim}
spde_cov <- inla.spde2.pcmatern(mesh = mesh_cov,
  alpha = 2,
  constr = TRUE, 
  prior.range = c(10, NA),
  prior.sigma = c(1, 0.01))
\end{verbatim}

\medskip
Note that in both models a sum-to-zero constraint has been included. This is to help with the identification of these exposure effects as there might be some confounding with the spatial effects.

\medskip
As stated above, when the exposure effect is considered as a typical linear effect, the value of the argument \texttt{model} is \texttt{"linear"}. However, when the effect is explicitly defined, the object where the effect has been saved has to be assigned to the argument \texttt{model}. In our case, we have saved the definition of the effect in \texttt{spde\_cov}, therefore, we have specified \texttt{model = spde\_cov}. 

\begin{figure}[h!]
\begin{minipage}{\textwidth}\vspace{-1.5 cm}
\begin{minipage}{0.45\textwidth}
\centering
\includegraphics[width = 6.85 cm, height = 6.85 cm]{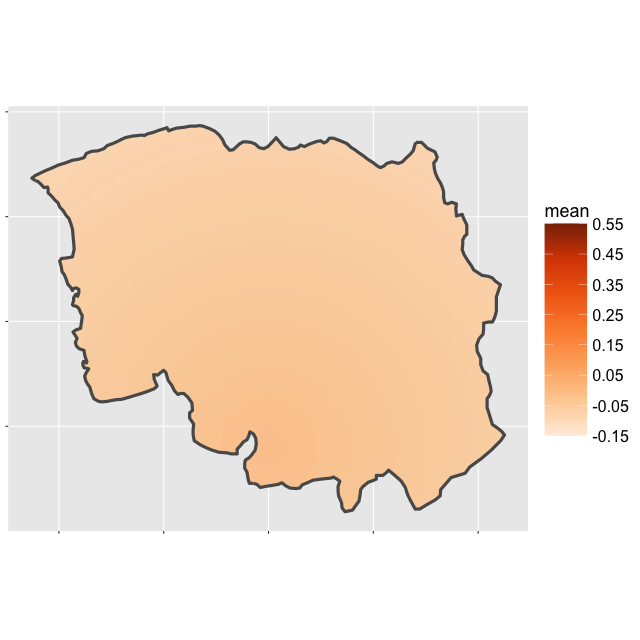}
\end{minipage}\hspace{0.75 cm}
\begin{minipage}{0.45\textwidth}
\centering
\includegraphics[width = 6.85 cm, height = 4.85 cm]{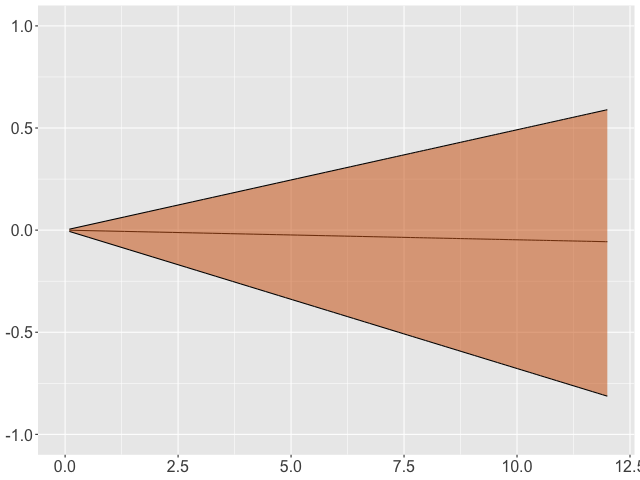}
\end{minipage} 
\end{minipage}

\begin{minipage}{\textwidth}\vspace{-1.5 cm}
\begin{minipage}{0.45\textwidth}
\centering
\includegraphics[width = 6.85 cm, height = 6.85 cm]{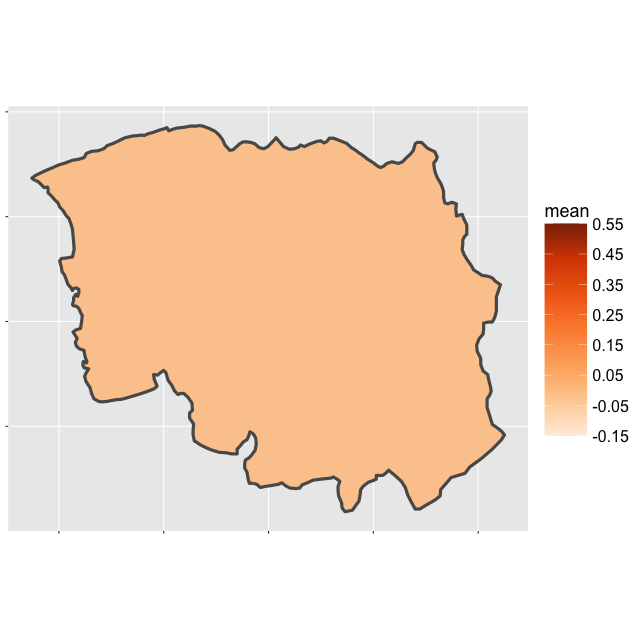}
\end{minipage}\hspace{0.75 cm}
\begin{minipage}{0.45\textwidth}
\centering
\includegraphics[width = 6.85 cm, height = 4.85 cm]{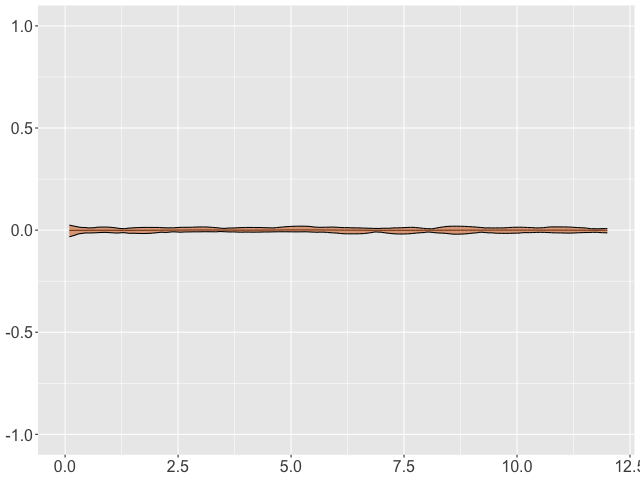}
\end{minipage}
\end{minipage}

\begin{minipage}{\textwidth}\vspace{-1.5 cm}
\begin{minipage}{0.45\textwidth}
\centering
\includegraphics[width = 6.85 cm, height = 6.85 cm]{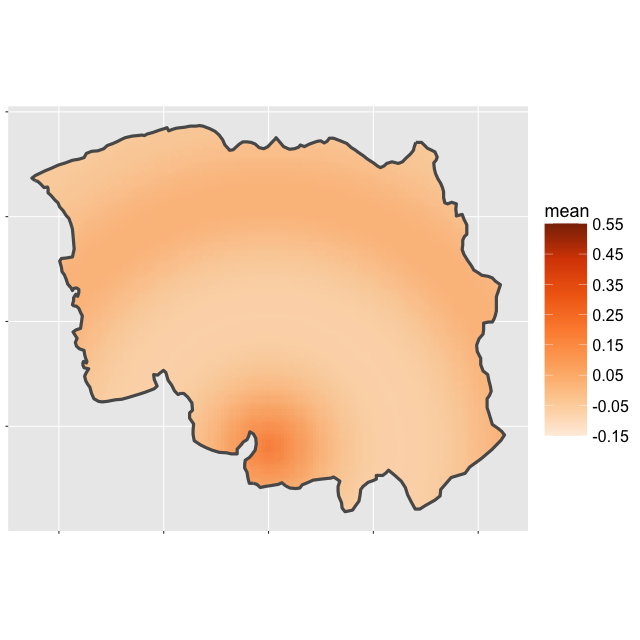}
\end{minipage}\hspace{0.75 cm}
\begin{minipage}{0.45\textwidth}
\centering
\includegraphics[width = 6.85 cm, height = 4.85 cm]{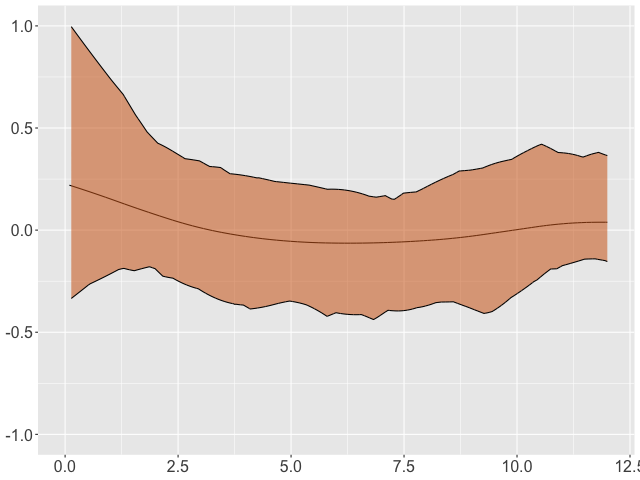}
\end{minipage}
\end{minipage}

\caption{The estimated posterior mean of the effect of the distance from the incinerator over the study region (left) and according to the distance to source (right) for linear effect (top), a SPDE of one dimension (center) and a random walk of second order (bottom).}
\label{fig4}
\end{figure}

\medskip
Regardless of the structure of the random effect used, the way this effect is introduced into the linear predictor is by setting argument \texttt{model} when defining the component:

\begin{verbatim}
cmp <- geometry ~ -1 + Inter.con(1) + Inter.lar(1) +
  sharedspde(geometry, model = pcmatern) +
  larspde(geometry, model = pcmatern) +
  dist(dist_terra, model = spde_cov)
\end{verbatim}

\medskip
This definition of the components can then be passed on to define the different likelihoods and fit the desired models. Figure~\ref{fig4} displays the estimated effect on the intensity of the larynx cancer cases according to distance to the incinerator.

\medskip
 \citet{Diggle1990splancs} reports a significant increase of cases in the vicinity of the incinerator. However, he also mentions that this is due to a cluster of four cases. This effect is clearly captured by the negative trend of the coefficient of the linear model (Figure \ref{fig4}, top-right plot) as well as the non-linear effect shown in the random walk smooth term (Figure \ref{fig4}, bottom-right plot). The SPDE model does not seem to capture the effect so clearly.
\citet{Diggle1990splancs} also conducts a sensitivity analysis by removing one or two cases from the small cluster of four, which leads to a decreasing evidence in the association. Our results illustrate the high uncertainty about the effect as shown by the wide credible intervals. Note that increasing the sample size to reduce the uncertainty is not an option in most spatial epidemiology studies.
Futhermore, a short simulation study has been included in \ref{app:simstudy} to assess that all three effects prposed here can effectively increased incidence around pollution sources when it is present.

\medskip
Computing times of the different models range from 24.71 seconds (univariate model for the larynx cancer cases) to 3.01 minutes (multivariate model with shared and specific components and no covariates). These times represent the duration required to execute the entire scripts (which include data loading, handling and visualization of results). Scripts were run on a Windows 10 LTSC Enterprise (1809) OS computer equipped with a 12th Gen Intel(R) Core(TM) i7-12700K processor running at 3.61GHz and 32GB of RAM.

\section{Discussion}
\label{sec:disc}

\medskip
The analysis of point patterns in spatial epidemiology often requires the
study of spatial risk variation as well as to assess potential risk
factors. In particular, the assessment of an increased risk in the vicinities
of putative pollution sources is of major interest. In this paper, we have
illustrated how log-Gaussian Cox processes can be widely adopted to tackle the
problems of estimation of the spatial risk variation and the assessment of the
effect of spatial and non-spatial risk factors.  In addition, Bayesian
inference about the required models have been conducted using the integrated
nested Laplace approximation. This allows the estimation of spatial random
effects based on Gaussian processes with a Matérn covariance, which can be efficiently estimated by means of the approximation based on stochastic partial differential equations. In addition, some guidelines have been stated and discussed in order to chose appropriate values for setting prior distributions on the hyperparameters of the (Mat\'ern covariance function) spatial effect. 

\medskip
The use of the proposed methods has been exemplified using a well-known
case-control dataset and the procedure to conduct the analysis has been
explained step by step. For this, the \texttt{inlabru} R package has been used
as it provides a simple way of defining and fitting the required models.
Our analysis considers the different models present in this paper and the results are consistent with the current literature. The complete analysis conducted
on the dataset can be
reproduced from the R code available
at \texttt{https://github.com/FranciscoPalmiPerales/MVPP-inlabru}.

\medskip
In conclusion, this paper has introduced an accessible, user-friendly, and efficient toolbox for analysing multivariate point patterns within an epidemiological context, without requiring expertise in statistics or programming. Specifically, the \texttt{inlabru} R package provides a straightforward and approachable framework for epidemiologists to analyse multivariate point pattern data.

\section{Acknowledgements}

This work has been supported by grant PID2019-106341GB-I00, funded by Ministerio de Econom\'ia y Competitividad (Spain), grant PID2022-136455NB-I00, funded by MCIN/AEI/10.13039/501100011033/FEDER, grant CIAICO/2022/165, funded by Dirección General de Ciencia e Investigación (Generalitat Valenciana) together with the European Regional Development Fund, grants SBPLY/17/180501/000491 and SBPLY/21/180501/000241, funded by Consejer\'ia de Educaci\'on, Cultura y Deportes (Junta de Comunidades de Castilla-La Mancha, Spain) and FEDER. F. Palm\'i-Perales has been supported by a Ph.D. scholarship awarded by the University of Castilla-La Mancha (Spain).

This work has also been Funded by the European Union. Views and opinions expressed are however those of the author(s) only and do not necessarily reflect those of the European Union or the European Climate, Infrastructure and Environment Executive Agency (CINEA). Neither the European Union nor the granting authority can be held responsible for them. 

\bibliographystyle{elsarticle-num-names} 
\bibliography{MultppInlabru}

\appendix

\section{Simulation study}
\label{app:simstudy}

In order to assess the validity of the methods presented in the paper a simple
simulation study has been included here. The aim is not to provide a thorough
exploration of the methods but to investigate wether the methods are able
to detect mild clustering. A common problem in spatial epidemiology studies 
is that the sample size cannot easily be increased.

A cluster of 5 points around the old incinerator has beeen added to the original data. \cite{Diggle1990splancs} reports an increased incidence in the original data created by a cluster of 4 points. Hence, clustered data around the source is scarce. The cluster of five points has been generated by adding five new points
with coordinates equal to that of the old incinerator plus some random perturbation. This is taken to be a Gaussian distribution with zero mean and standard deviation of 0.5 km (added independently to each coordinate). This means that now there are 978 controls and 63 cases.

\begin{figure}[h!]
\centering
\includegraphics[scale=0.3]{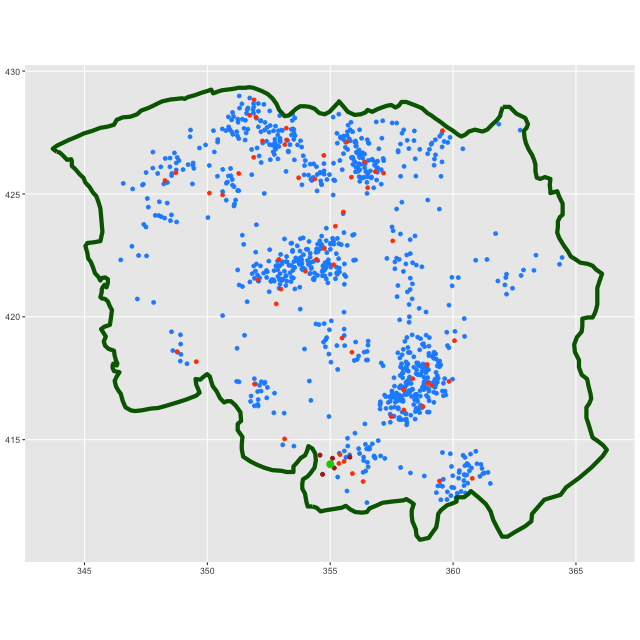}
\includegraphics[scale=0.3]{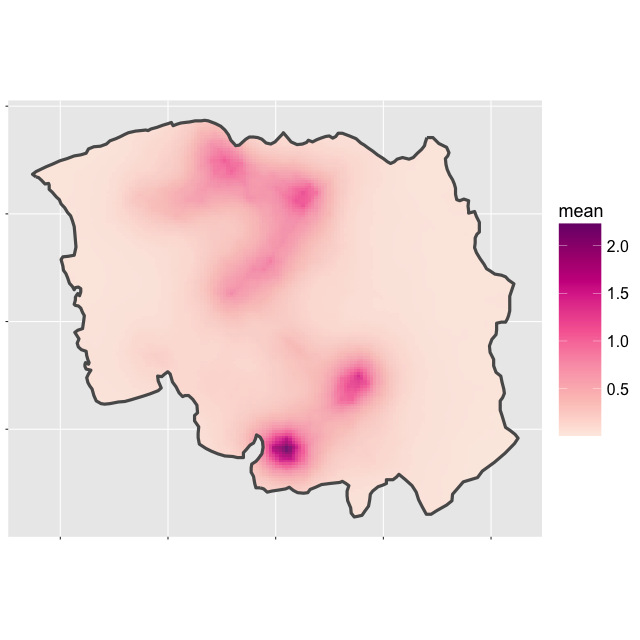} 
\caption{Location of cases, controls and the old incinerator (left plot) and intensity estiamted from the new larynx cancer cases. The new five clustered cases have been coloured in brown in the left plot.}
\label{app:data}
\end{figure}

Figure~\ref{app:data} shows a map with the different points (left plot) and the intensity estimated from the cases of larynx cancer (that now includes the
cluster of 5 points).

\begin{figure}[h!]
\centering
\includegraphics[scale=0.3]{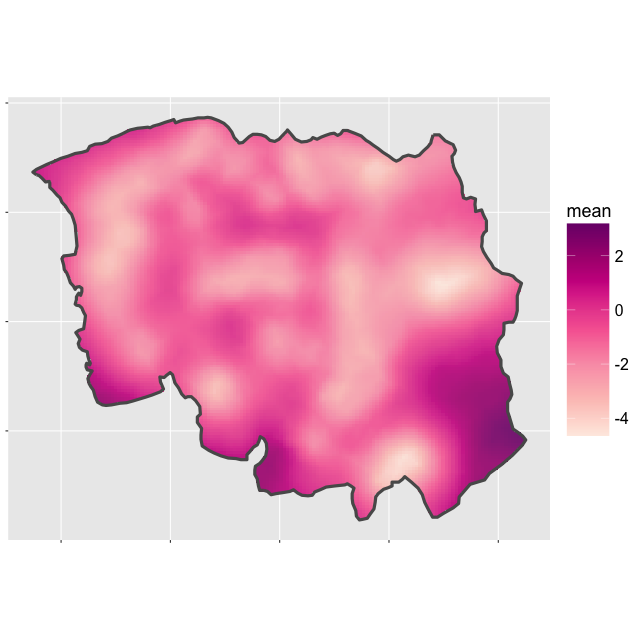}
\caption{Difference of the spatial random effects of the intensities of cases and controls.}
\label{app:ratio}
\end{figure}

Figure~\ref{app:ratio} illustrates the spatial distribution of the log-ratio between the (rescaled) intensities, i.e., it represents the difference between
the spatial effect of the cases and that of the controls. A value of zero represents similar spatial variation while values larger than zero represent clustering of the cases. 

\begin{figure}[h!]
\begin{minipage}{\textwidth}\vspace{-1.5 cm}
\begin{minipage}{0.45\textwidth}
\centering
\includegraphics[width = 6.85 cm, height = 6.85 cm]{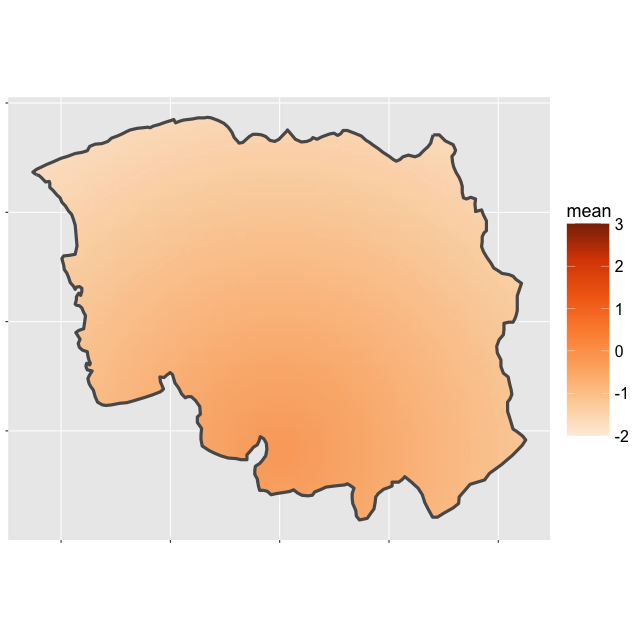}
\end{minipage}\hspace{0.75 cm}
\begin{minipage}{0.45\textwidth}
\centering
\includegraphics[width = 6.85 cm, height = 4.85 cm]{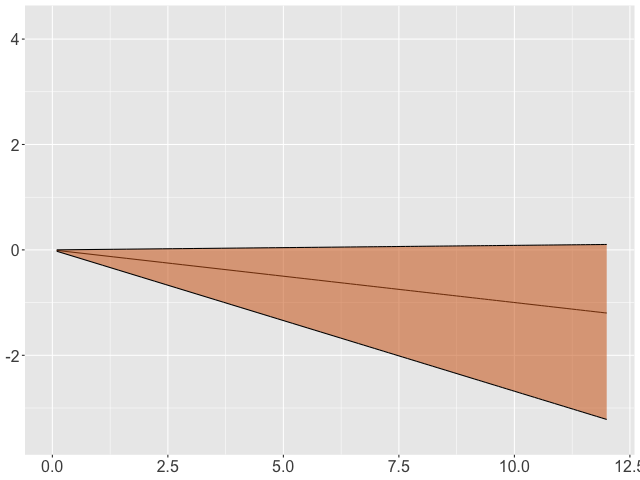}
\end{minipage}
\end{minipage}

\begin{minipage}{\textwidth}\vspace{-1.5 cm}
\begin{minipage}{0.45\textwidth}
\centering
\includegraphics[width = 6.85 cm, height = 6.85 cm]{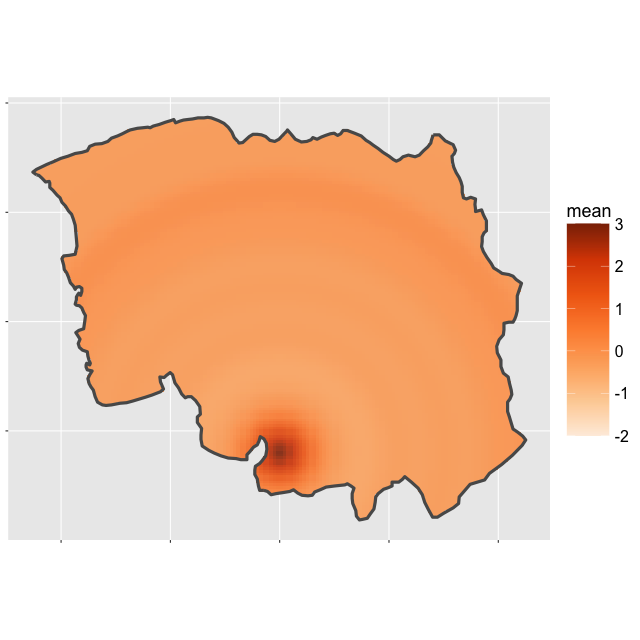}
\end{minipage}\hspace{0.75 cm}
\begin{minipage}{0.45\textwidth}
\centering
\includegraphics[width = 6.85 cm, height = 4.85 cm]{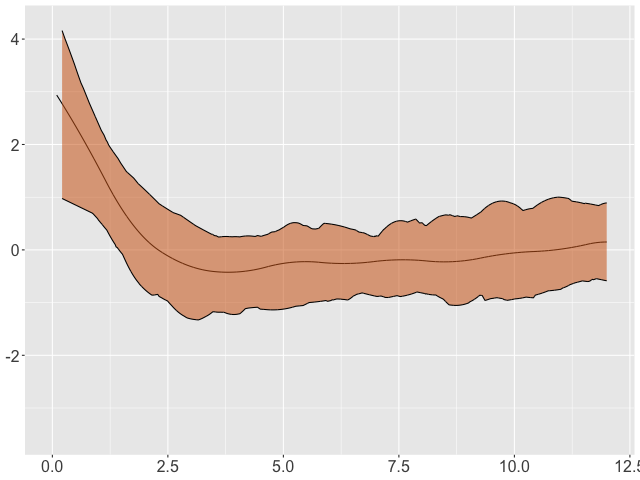}
\end{minipage}
\end{minipage}

\begin{minipage}{\textwidth}\vspace{-1.5 cm}
\begin{minipage}{0.45\textwidth}
\centering
\includegraphics[width = 6.85 cm, height = 6.85 cm]{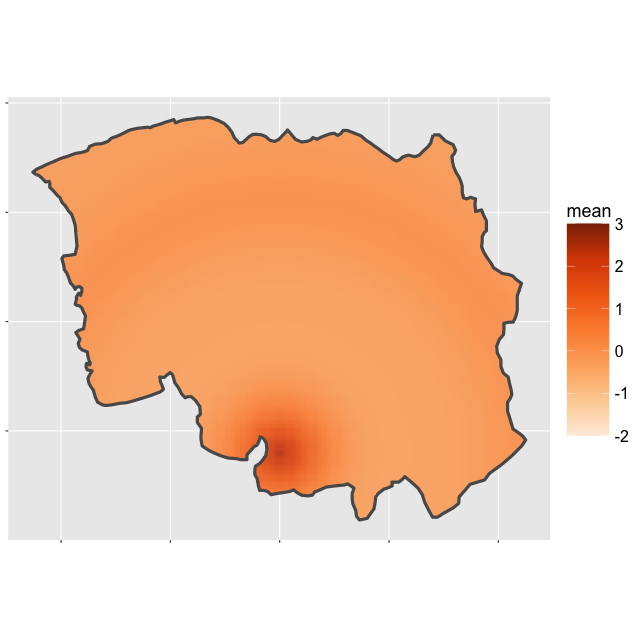}
\end{minipage}\hspace{0.75 cm}
\begin{minipage}{0.45\textwidth}
\centering
\includegraphics[width = 6.85 cm, height = 4.85 cm]{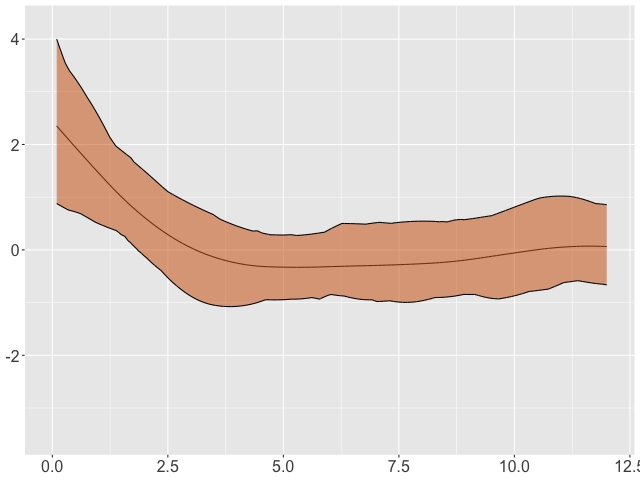}
\end{minipage}
\end{minipage}

\caption{The estimated posterior mean of the effect of the distance from the incinerator over the study region (left) and according to the distance to source (right) for linear effect (top), a SPDE of one dimension (center) and a random walk of second order (bottom).}
\label{app:covariate}
\end{figure}

Finally, Figure~\ref{app:covariate} shows the estimates of the different effects
to assess clustering around the old incinerator. As it can be seen, all three effects capture the increase of cases around the pollution source. It is worth mentioning that both the SPDE and random walk of order 2 provide positive estimates of the effect at short distances with credible intervals that leave zero below.

Hence, it is clear that the models proposed in this paper are useful when modeling spatial variation of case-control point patterns to assess spatial variation and clustering around putative pllution sources.

\end{document}